\documentclass{iaus}
\usepackage{graphicx}

\title[High-ionization recombination lines in NGC 6302. ] 
{NGC 6302: high-ionization permitted lines. {\it Applying X-SSN synthesis to VLT-UVES spectra}}

\author[D. P\'equignot, C. Morisset \& S. Casassus]   
{Daniel P\'equignot$^{1}$,
 Christophe Morisset$^{2, 3}$,
\and Simon Casassus$^{4}$}

\affiliation{
$^1$LUTh, Observatoire de Paris-Meudon, France {\tt Daniel.Pequignot@obspm.fr} \\[\affilskip]
$^2$Instituto de Astronom\'{\i}a, Universidad Nacional Aut\'onoma de M\'exico, Mexico. \\[\affilskip]
$^3$Instituto de Astrof\'{\i}sica de Can\'arias, La Laguna, Tenerife, Spain. \\[\affilskip]
$^4$Universidad de Chile, Departamento de Astronom\'{\i}a, Chile.}

\pubyear{2011}
\volume{283}  
\pagerange{}
\jname{Planetary Nebulae: An Eye to the Future}
\editors{A. Manchado, \& L. Stanghellini, eds.}
\begin{document}

\maketitle

\begin{abstract}
A preliminary VLT-UVES spectrum of NGC~6302 (Casassus et al. 2002, MN), which hosts one of the hottest PN nuclei known (Teff $\sim$ 220000\,K; Wright et al. 2011, MN), has been recently analysed by means of X-SSN, a spectrum synthesis code for nebulae (Morisset and P\'equignot). Permitted recombination lines from highly-ionized species are detected/identified for the first time in a PN, and some of them probably for the first time in (Astro)Physics. The need for a homogeneous, high signal-to-noise UVES spectrum for NGC~6302 is advocated. 
\end{abstract}


VLT-UVES allows to detect emission lines at resolution 50,000 down to a few 10$^{-6}$ I$_{H\beta}$ flux in bright PNe such as NGC~6302, raising the problem of properly extracting all pieces of information from increasingly complex spectra. Spectrum synthesis is the way to master complexity and record useful findings to be cross-checked with other data. X-SSN computes and works with line sets {\sl specifically relevant to nebulae}, each of them controlled by the intensity and intrinsic profile of a reference line, and accounts for all of the modulations (reddening, DIBs, interstellar lines, telluric lines, and final convolution with instrumental profile) required to match the synthetic spectrum to the observations. A process of convergence bringing the synthetic spectrum into coincidence with the observed spectrum is taking over from the traditional measurement of individual lines. Instead of just emphasizing wavelength coincidence, spectral synthesis considerably helps line identification thanks to relative line intensities, allowing also for objective and systematic line deblending. 

Establishing a deep template spectrum for high-excitation nebulae is beneficial to several fields of Astrophysics (AGN NLR, novae). Also, by recording the many 'anonymous' weak lines from abundant elements, which generate a highly detrimental 'pseudo-noise', template spectra contribute to create conditions to access subtle, yet potentially important new spectral signatures in nebulae. Conversely, UVES observations of NGC~6302 can bring necessary inputs to X-SSN, whose ultimate aim is to provide a new service. 
Atomic Physics remains a foundation of Astrophysics. Despite the relative coarseness and incompleteness of the spectrum (obtained during the test phase of UVES with imperfect filters and relatively short exposures), the few examples commented below amply demonstrate that, provided that a sufficiently deep spectrum can be secured, NGC~6302 with UVES and X-SSN offers a rare opportunity to implement the interface of these sciences, while providing original ionic abundance diagnostics: six well-documented ionization stages for both oxygen and neon appears to be within reach of observation, then suppressing missing ionization stages, constraining detailed photoionization models, and shedding new light on the so-called ’Abundance Discrepancy Factor’ problem. 

\begin{figure}
\begin{center}
 \includegraphics[width=13.4cm]{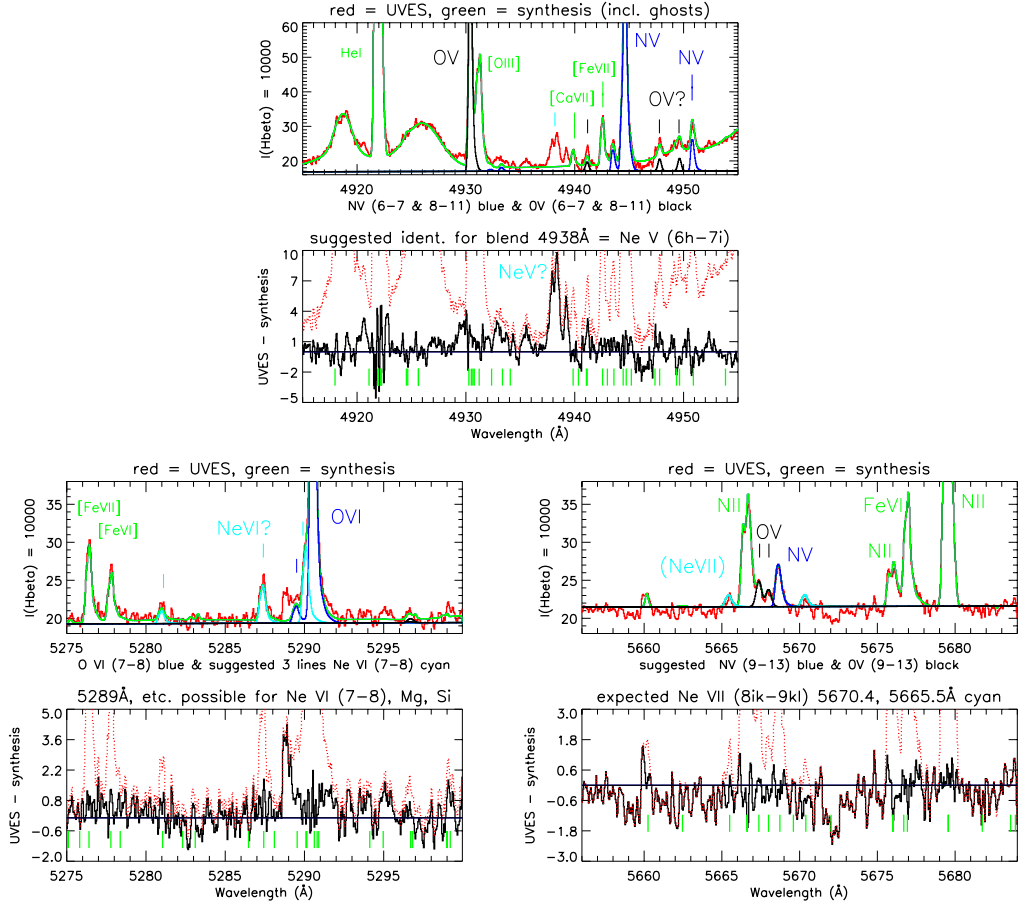} 
 \caption{
{\bf Fig.~a (top):} Example of {\bf X-SSN} spectral synthesis. The lower panel shows the residuals (amplified), and highlights the suggested Ne\,{\sc v} (6h-7i), which is not included in the fit. The uncertain wavelengths for the two Ne\,{\sc v} (6h-7i) and three O\,{\sc v} (8-11) lines are tentatively determined here from the observed spectrum (and considering their theoretical intensities). New UVES spectra and atomic data can provide reliable identifications. Over 4915-55\AA, GDWH (Groves et al. 2002, PASA)
 identified two lines (4922 \& 4931) and recorded two more lines, that they left unidentified (4938 \& 4945). Applying spectral synthesis to the high resolution data of UVES is a powerful tool to improve line wavelengths. N\,{\sc v} (6-7) 4944.5 was not identified by GDWH because they relied on slightly wrong wavelengths and ignored intensities. Quite probably, {\sl hundreds of atomic levels can be checked in this way}. At their resolution, GDWH also missed the strong O\,{\sc v} (6h-7i) 4930\AA. Ne\,{\sc v} recombination lines have {\sl never been identified before} in any nebula. 
{\bf Fig.~b (lower left):} The correct identification for $\lambda$5290\AA\ is not [Fe\,{\sc vi}] (GDWH), but O\,{\sc vi}(7-8): to our knowledge, this is the {\sl first identification} of a nebular (not Wolf-Rayet) O\,{\sc vi} recombination line in a PN. Other O\,{\sc vi} lines are detected in the UVES spectrum with coherent intensities. Interestingly, there is a reasonable suspicion that Ne\,{\sc vi}, {\sl never seen before} in any nebula, is detected by means of three lines. A new spectrum is essential to ascertain the presence of the lines and give a status to other ill-defined features, notably $\lambda$5289\AA, which have no identifications so far.
{\bf Fig.~c (lower right):} Not only the strongest, but a full set of emission lines in a broad wavelength range must be considered at sufficient resolution to check and establish useful atomic models for nebulae. Here, it appears that, relative to the quasi-hydrogenic N\,{\sc v}(9-13) line, O\,{\sc v}(9-13) must be slightly shifted and split into at least two components in order to fit in the observed spectrum. New data can make the decision. A higher S/N could also lead to Ne\,{\sc vii} detection (2 lines). Acheiving a limit of one tenth the N\,{\sc v}(9-13) flux, some 30 distinct N\,{\sc v}, O\,{\sc vi} and O\,{\sc iv}, 40 N\,{\sc iii}, 60 O\,{\sc v}, and 90 N\,{\sc iv} radiative recombination lines should be detectable, not including a number of small-$nl$ and di-electronic lines.
} 
   \label{figs}
\end{center}
\end{figure}


\end{document}